\documentclass[10pt,journal,compsoc]{IEEEtran}

\usepackage{multirow}
\usepackage{graphicx}
\usepackage{algorithm}
\usepackage{algorithmic}
\usepackage{amsfonts}
\usepackage{subfigure} 
\usepackage{verbatim}
\usepackage{epsfig}

\ifCLASSOPTIONcompsoc
\else
\fi
\ifCLASSINFOpdf
\else
\fi
\usepackage[cmex10]{amsmath}
\begin{document}
%
\title{Automated Analysis and Prediction of Job Interview Performance}
%
%
%
%

\author{Iftekhar~Naim,~\IEEEmembership{Student Member,~IEEE,}
        M. Iftekhar~Tanveer,~\IEEEmembership{Student Member,~IEEE,}
        Daniel~Gildea
        and~Mohammed~(Ehsan)~Hoque,~\IEEEmembership{Member,~IEEE}
\thanks{}}

\IEEEcompsoctitleabstractindextext{%
\begin{abstract}
We present a computational framework for automatically quantifying verbal and nonverbal behaviors in the context of job interviews. The proposed framework is trained by analyzing the videos of 138 interview sessions with 69 internship-seeking undergraduates at the Massachusetts Institute of Technology (MIT). Our automated analysis includes facial expressions (e.g., smiles, head gestures, facial tracking points), language (e.g., word counts, topic modeling), and prosodic information (e.g., pitch, intonation, and pauses) of the interviewees. The ground truth labels are derived by taking a weighted average over the ratings of 9 independent judges. Our framework can automatically predict the ratings for interview traits such as excitement, friendliness, and engagement with correlation coefficients of 0.75 or higher, and can quantify the relative importance of prosody, language, and facial expressions. By analyzing the relative feature weights learned by the regression models, our framework recommends to speak more fluently, use less filler words, speak as ``we" (vs. ``I"), use more unique words, and smile more. We also find that the students who were rated highly while answering the first interview question were also rated highly overall (i.e., first impression matters). Finally, our MIT Interview dataset will be made available to other researchers to further validate and expand our findings.
\end{abstract}
\begin{IEEEkeywords}
Nonverbal Behavior Prediction, Job Interviews, Multimodal Interactions, Regression.
\end{IEEEkeywords}}

\maketitle

\IEEEdisplaynotcompsoctitleabstractindextext

%
\IEEEpeerreviewmaketitle

\section{Introduction}
Analysis of non-verbal behavior to predict the outcome of a social interaction has been studied for many years in different domains, with predictions ranging from marriage stability based on interactions between newlywed couples~\cite{gottman77,gottman-marriage03}, to patient satisfaction based on doctor-patient interaction~\cite{hall09}, to teacher evaluation by analyzing classroom interactions between a teacher and the students~\cite{ambady93}.
However, many of these prediction frameworks were based on manually labeled behavioral patterns by trained coders, according to carefully designed coding schemes. 
Manual labeling of nonverbal behaviors is laborious and time consuming, and therefore often does not scale with large amounts of data.
As a scalable alternative, several automated prediction frameworks have been proposed based on low-level behavioral features, automatically extracted from larger datasets.
Due to the challenges of collecting and analyzing multimodal data, most of these automated methods focused on a single modality of interaction~\cite{Curhan07,Sandbach12,Castellano07,Soman-ICASSP10}.
In this paper, we address the challenge of automated understanding of multimodal human interactions, including facial expression, prosody, and language. 
We focus on predicting social interactions in the context of job interviews for college students, which is an exciting and relatively less explored domain.
\begin{figure}
\label{fig:system}
\centering
\includegraphics[width=0.40\textwidth]{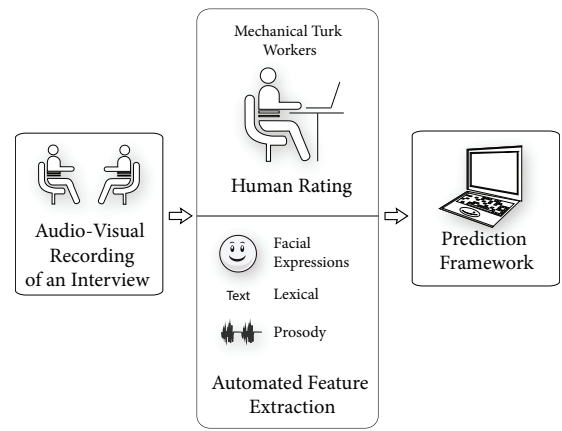}
\caption{Framework of Analysis. Mechanical Turk workers rated interviewee performance by watching videos of job interviews. Various features were extracted from those videos. A framework was built to predict Turker's rating and to gain insight into the characteristics of a good interview.}
\end{figure}

Job interviews are ubiquitous and play inevitable and important roles in our life and career. 
Over many years, social psychologists and career coaches have accumulated knowledge and guidelines for success in job interviews~\cite{Huffcutt-jap01,Posthuma-PerPsy02,Macan-HRMR09}. 
Studies in social psychology have shown that smiling, using a louder voice, and maintaining eye contact contribute positively to our interpersonal communications~\cite{Huffcutt-jap01,Macan-HRMR09}. 
These guidelines are largely based on intuition, experience, and studies involving manual encoding of nonverbal behaviors on a limited amount of data~\cite{Huffcutt-jap01}. 
Automated data-driven quantification of both verbal and non-verbal behaviors simultaneously has not been explored in the context of job interviews.
In this paper, we aim to quantify the determinants of a successful job interview using a computational prediction framework based on automatically extracted features, which takes both verbal speech and non-verbal behaviors into account.

Imagine the following scenario in which two students, John and Matt, were individually asked to discuss their leadership skills in a job interview. John responded with the following: 
\begin{quote}
``\emph{One semester ago, I was part of a team of ten students [stated in a loud and clear voice]. 
We worked together to build an autonomous playing robot. 
I led the team by showing how to program the robot. 
The students did a wonderful job [conveyed excitement with tone]!
In ten weeks, we made the robot play soccer. It was a lot of fun. [concluded with a smile]''. }
\end{quote}
Matt responded with the following:
\begin{quote}
 ``\emph{Umm ...  [paused for 2 seconds] Last semester I led a group in a class project on robot programming. It was a totally crazy experience. The students did almost nothing until the last moment. ... Umm ... Basically, I had to intervene at that point and led them to work hard. Eventually, this project was completed successfully. [looked away from the interviewer]}". 
\end{quote}
Who do you think received higher ratings?

Most would agree that the first interviewee, John, provided the more enthusiastic and engaging answer. 
We can easily interpret the meaning of our verbal and nonverbal behavior during face-to-face interactions. 
However, we often cannot quantify how the combination of these behaviors affects our interpersonal communications. 
Previous research~\cite{Kapoor05} shows that the style of speaking, prosody, facial expression, and language reflect valuable information about one's personality and mental states. 
Understanding the relative influence of these individual modalities can provide crucial insight regarding job interviews.

In this paper, we attempt to answer the following research questions by analyzing the audio-visual recordings of 138 interview sessions with 69 individuals:
\begin{itemize}
\item Can we automatically quantify verbal and nonverbal behavior, and assess their role in the overall rating of job interviews? 

\item Can we build a computational framework that can automatically predict the overall rating of a job interview given the audio-visual recordings?

\item Can we infer the relative importance of language, facial expressions, and prosody (intonation)?

\item Can we make automated recommendations on improving social traits such as excitement, friendliness, and engagement in the context of a job interview?
\end{itemize}

To answer these research questions, we designed and implemented an automated prediction framework for quantifying the ratings of job interviews, given the audio-visual recordings. 
The proposed prediction framework (Figure~\ref{fig:system}) automatically extracts a diverse set of multimodal features (lexical, facial, and prosodic), and quantifies the overall interview performance, the likelihood of getting hired, and 14 other social traits relevant to the job interview process. 
Our system is capable of predicting the overall rating of a job interview with a correlation coefficient $r > 0.65$ and AUC = 0.81 (baseline 0.50) on average. 
We can also predict different social traits such as engagement, excitement, and friendliness with even higher accuracy ($r \geq 0.75$, AUC $> 0.85$). 
Furthermore, we investigate the relative weights of the individual verbal and non-verbal features learned by our regression models, and quantify their relative importance in the context of job interviews. 
Our prediction model can be integrated with the existing automated interview coaching systems, such as MACH~\cite{Hoque-ubicomp13}, to provide more intelligent and quantitative feedback. 
The interview questions asked in our training dataset are chosen to be independent of any job specifications or skill requirements. 
Therefore, the ratings predicted by our model are based on social and behavioral skills only, and they may differ from a hiring manager's opinion, given a specific job.

Parts of the research included in this article have been presented in~\cite{Naim-FG15}.
In this article, we present an improved system by including additional facial features and provide more comprehensive results and analysis.
The remaining structure of the article follows. In Section~\ref{section:background}, we discuss the background research on automated quantification of multimodal nonverbal behaviors. Section~\ref{section:data} describes the interview dataset and the data annotation process via Mechanical Turk. 
A detailed discussion of the proposed computational framework, feature extraction, and automated prediction is presented in Section~\ref{section:methods}. 
We present our detailed results in Section~\ref{section:results}. Finally, we conclude with our findings and discuss our future work in Section~\ref{section:discussions}.
\section{Background Research}
\label{section:background}
In this section, we discuss existing relevant work on nonverbal behavior prediction using automatically extracted features. 
We particularly focus on the social cues that have been shown to be relevant to job interviews and face-to-face interactions~\cite{Huffcutt-jap01}. 
We also discuss previous research on automated conversational systems for job interviews, which is one of the potential applications we envision for the proposed prediction framework.

\subsection{Nonverbal Behavior Recognition}
Nonverbal behaviors are subtle, fleeting, subjective, and sometimes even contradictory. 
Even a simple facial expression such as a smile can have different meanings, e.g., delight, rapport, sarcasm, and even frustration~\cite{Hoque-AC12}. 
Edward Sapir, in 1927, referred to non-verbal behavior as ``an elaborate and secret code that is written nowhere, known by none, but understood by all''~\cite{sapir1985selected}.
Despite years of research, nonverbal behavior prediction remains a challenging problem. 
Gottman et al.~\cite{gottman77,gottman-marriage03} studied verbal and non-verbal interactions between newlywed couples and developed mathematical models to predict marriage stability and chances of divorce. 
For example, they found that the greatest predictor of divorce is contempt, which must be avoided for a successful marriage. 
Hall et al.~\cite{hall09} studied the non-verbal cues in doctor-patient interaction and showed that doctors who are more sensitive to nonverbal skills received higher ratings of service during patient visits.
Ambady et al.~\cite{ambady93} studied the interactions of teachers with students in a classroom and proposed a framework for predicting teachers' evaluations based on short clips of interactions.
However, these prediction frameworks were based on manually labeled behavioral patterns.
Manually labeling non-verbal behaviors is laborious and time consuming, and is often not scalable to large amounts of data.

To allow for the analysis of larger datasets of social interactions, several automated prediction frameworks have been proposed. 
Due to the challenges of collecting and analyzing multimodal data, most of the existing automated prediction frameworks focus on a single behavioral modality, such as prosody~\cite{Soman-ICASSP10,Frick-PsyBultn85,Zechner09}, facial expression~\cite{Sandbach12}, gesture~\cite{Castellano07}, and word usage pattern~\cite{Tausczik-jlsp10}. 
Analysis based on a single modality is likely to overlook many critical non-verbal behaviors, and hence there has been a growing interest in analyzing social behaviors in more than a single modality.

Ranganath et al.~\cite{Ranganath-emnlp09,Ranganath-CSL13} studied social interactions in speed-dates using a combination of prosodic and linguistic features. 
The analysis is based on the SpeedDate corpus, a spoken corpus of approximately 1000 4-min-speed-dates, where each participant rated his/her date in terms of four different conversational styles (awkwardness, assertiveness, flirtatiousness, and friendliness) on a ten point Likert scale.
Given the speech data, Ranganath et al. proposed a computational framework for predicting these four conversational styles using prosodic and linguistic features only, while ignoring facial expressions.
Stark et al.~\cite{stark14} were able to reliably predict the nature of a telephone conversation (business versus personal, familiar versus unfamiliar) using the lexical and prosodic features extracted from as few as 30 words of speech at the beginning of the conversation.
Kapoor et al.~\cite{Kapoor05} and Pianesi et al.~\cite{Pianesi-ICMI08} proposed systems to recognize different social and personality traits by exploiting only prosody and visual features. 
Sanchez et al.~\cite{Sanchez-13} proposed a system for predicting eleven different social moods (e.g., surprise, anger, happiness) from YouTube video monologues, which consist of different social dynamics than face to face interactions. 

Perhaps the most relevant, Nguyen et al.~\cite{Nguyen2014} proposed a computational framework to predict the hiring decision using nonverbal behavioral cues extracted from a dataset of 62 interview videos. 
Nguyen et al. considered only nonverbal cues, and did not include verbal content in their analysis. 
Our work extends the current state-of-the-art and generates new knowledge by incorporating three different modalities (prosody, language, and facial expressions), and fifteen different social traits (e.g., friendliness, excitement, engagement), and quantifies the interplay and relative influences of these different modalities for each of the different social traits. 
Furthermore, by analyzing the relative feature weights learned by our regression models, we obtain valuable insights about behaviors that are recommended for success in job interviews (Section~\ref{section:recommendation}).

\subsection{Social Coaching for Job Interviews}
Several automated systems have been proposed for coaching the necessary social skills to succeed in job interviews~\cite{Hoque-ubicomp13,Anderson-ACE13,Baur-HBU13}. 
Hoque et al.~\cite{Hoque-ubicomp13} developed MACH (My Automated Conversation coacH), which allows users to improve social skills by interacting with a virtual agent. 
The MACH system records videos of the user using a webcam and a microphone, and provides feedback regarding several low level behavioral patterns, e.g., average smile intensity, pause duration, speaking rate, pitch variation, etc. 

Anderson et al.~\cite{Anderson-ACE13} proposed an interview coaching system, TARDIS, which presents the training interactions as a scenario-based ``serious game''. 
The TARDIS framework incorporates a sub-module named NovA (NonVerbal behavior Analyzer)~\cite{Baur-HBU13} that can recognize several lower level social cues: \emph{hands-to-face}, \emph{looking away}, \emph{postures}, \emph{leaning forward/backward}, \emph{gesticulation}, \emph{voice activity}, \emph{smiles}, and \emph{laughter}. 
Using videos that are manually annotated with these ground truth social cues, NovA trains a Bayesian Network that can infer higher-level mental traits (e.g., stressed, focused, engaged, etc.). Automated prediction of higher-level traits remains part of their future work.

Our framework (1) quantifies the relative influences  of different low level features on the interview outcome, (2)  learns regression models to predict interview ratings and the likelihood of hiring using automatically extracted features, and (3) predicts several other high-level personality traits such as engagement, friendliness, and excitement. One of our objectives is to extend the existing automated conversation systems by providing feedback on the overall interview performance and additional high-level personality traits. 
\section{Dataset Description}
\label{section:data}
We used the \emph{MIT Interview Dataset}~\cite{Hoque-ubicomp13}, which consists of 138 audio-visual recordings of mock interviews with internship-seeking students from Massachusetts Institute of Technology (MIT). 
The total duration of our interview videos is nearly 10.5 hours (on average, 4.7 minutes per interview, for 138 interview videos).
To our knowledge, this is the largest collection of interview videos conducted by professional counselors under realistic settings.
The following sections provide a brief description of the data collection and ground truth labeling.

\subsection{Data Collection}
\begin{figure}
\centering
\includegraphics[width=0.4\textwidth]{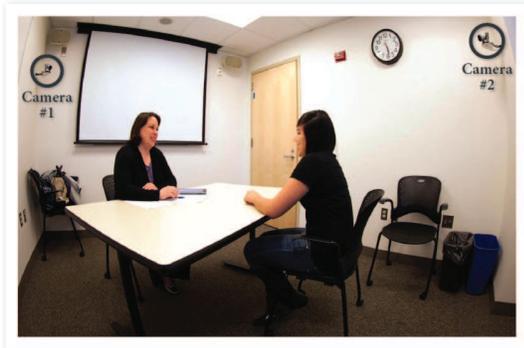}
\caption{The experimental setup for collecting audio-visual recordings of the mock interviews. Camera \#1 recorded the video and audio of the interviewee, while Camera \#2 recorded the interviewer.}
\label{fig:interviewSetup}
\end{figure}

\subsubsection{Study Setup}
The mock interviews were conducted in a room equipped with a desk, two chairs, and two wall-mounted cameras, as shown in Figure~\ref{fig:interviewSetup}. 
The two cameras with microphones were used to capture the facial expressions and the audio conversations during the interview.

\subsubsection{Participants}
Initially, 90 MIT juniors participated in the mock interviews. 
All participants were native English speakers. 
The interviews were conducted by two professional MIT career counselors  who had over five years of experience. For each participant, two rounds of mock interviews were conducted: before and after interview intervention. For the details of interview intervention, please see~\cite{Hoque-ubicomp13}.  
Each individual received \$50 for participating. 
Furthermore, as an incentive for the participants, we promised to forward the resume of the top 5\% candidates to several sponsor organizations (Deloitte, IDEO, and Intuit) for consideration for summer internships. 
We chose sponsor organizations which are not directly tied to any specific major.
After the data collection, 69 (26 male, 43 female) of the 90 initial participants permitted the use of their video recordings for research purposes.

\subsubsection{Procedure}
During each interview session, the counselor asked interviewees five different questions, which were recommended by the MIT Career Services. These five questions were presented in the following order by the counselors to the participants:
\begin{quote}
\emph{Q1. So please tell me about yourself.}\\
\emph{Q2. Tell me about a time when you demonstrated  leadership.}\\
\emph{Q3. Tell me about a time when you were working  with a team and faced a challenge. How did you overcome the problem?}\\
\emph{Q4. What is one of your weaknesses and how do you plan to overcome it?}\\
\emph{Q5. Now, why do you think we should hire you?} \\
\end{quote}
\vspace{-8pt}

\begin{table}
  \renewcommand{\arraystretch}{1.2}
\footnotesize
\caption{List of questions asked to Mechanical Turk workers. First two questions are related to interviewee performances. Others are on various traits of their behavior}
  \begin{tabular}{||  l | l ||  }  \hline \hline
	Traits  	& Description  \\  \hline \hline
	Overall Rating 	& The overall performance rating. \\
	Recommend Hiring & How likely is he to be hired?\\
	Engagement	& Did he use engaging tone? \\
	Excitement 	& Did he seem excited?\\
	Eye Contact 	& Did he maintain proper eye contact?\\
	Smile		 	& Did he smile appropriately?\\
	Friendliness 	& Did he seem friendly?\\
	Speaking Rate 	& Did he maintain a good speaking rate?\\
	No Fillers		 & Did he use too many filler words?\\
				&  (1 = too many, 7 = no filler words)\\
	Paused		 & Did he pause appropriately?\\
	Authentic		 & Did he seem authentic?\\
	Calm			 & Did he appear calm?\\
	Focused	 	& Did he seem focused?\\
	Structured Answers	 	& Were his answers structured?\\
	Not Stressed	 & Was he stressed?\\
				&  (1 = too stressed, 7 = not stressed)\\
	Not Awkward	 & Did he seem awkward?\\
				&  (1 = too awkward, 7 = not awkward)\\  \hline  \hline
  \end{tabular}
\label{table:traits}
\end{table}
\vspace{-5pt}

No job description was given to the interviewees. 
The five questions were chosen to assess the interviewee's behavioral and social skills. 
The interviewers rated the performances of the interviewees by answering 16 assessment questions on a seven point Likert scale. We list these questions in Table \ref{table:traits}. These questions to the interviewers were selected to evaluate the overall performance and behavioral traits of the interviewees. The first two questions -- ``Overall Rating'' and ``Recommend Hiring'' - represent the overall performance. The remaining questions have been selected to evaluate several high-level behavioral dimensions such as warmth (e.g., ``friendliness", ``smiling"), presence (e.g., ``engagement", ``excitement", ``focused"), competence (e.g. speaking rate), and content (e.g., ``structured").

\subsection{Data Labeling}
The subjective nature of human judgment makes it difficult to collect ground truth for interview ratings. 
Due to the nature of the experiment, the counselors interacted with each interviewee twice - before and after intervention, and provided feedback after each session. 
The process of feedback and the way the interviewees responded to the feedback may have had an influence on the counselor's ratings. In order to remove the bias introduced by the interaction, we used Amazon Mechanical Turk workers to rate the interview performance. 
The Mechanical Turkers used the same questionnaire to assess the ratings as listed in Table~\ref{table:traits}. 
Apart from being less affected by bias, the Mechanical Turk workers could pause and replay the video, allowing them to rate more thoroughly. 
The Turkers' ratings are more likely to be similar to ``audience'' ratings, as opposed to ``expert ratings".

In order to collect ground truth ratings for interviewee performances,  we first selected 10 Turkers out of 25, based on how well they agreed with the career counselors on the five control videos. 
Out of these 10 selected Turkers, one did not finish all the rating tasks, leaving us with 9 ratings per video. 
We automatically estimated the quality of individual workers using an EM-style optimization algorithm, and estimated a weighted average of their scores as the ground truth ratings, which were used in our prediction framework.

One of our objectives was to model the temporal relationships among the individual interview questions and the overall ratings. 
To accomplish this, we performed a second phase of labeling.
We hired a different set of 5 Turkers for rating the performances of an interviewee in each of the five interview question separately. 
This was done by splitting each interview video into five different segments, where each segment corresponds to one of the interview questions. 
The video segments were shuffled so that each Turker would rate the segments in a random order. 
These per-question ratings were used only to analyze the temporal variation in the ratings and measure how the temporal order of the questions correlates with the ratings for entire interview. 
\section{Prediction Framework}
\label{section:methods}
For the prediction framework, we automatically extracted various features from the videos of the interviews. Then we trained two regression algorithms - SVM and LASSO. The objective of this training is twofold: first, to predict the Turker's ratings on the overall performance and each behavioral trait, and second, to quantify and gain meaningful insights on the relative importance of each modality and the interplay among them.  

\subsection{Feature Extraction}
We collected three types of features for each interview video: (1) prosodic features, (2) lexical features, and (3) facial features. We selected these features to reflect the behaviors that have been shown to be relevant in job interviews (e.g., smile, intonation, language content, etc.)~\cite{Huffcutt-jap01}, and also based on the past literature on automated social behavior recognition~\cite{Sanchez-13,Ranganath-emnlp09,Ranganath-CSL13,Zechner09}. 
For extracting reliable lexical features, we chose not to use automated speech recognition. Instead, we transcribed the videos by hiring Amazon Mechanical Turk workers, who were specifically instructed to include filler and disfluency words (e.g., ``uh'', ``umm'', ``like'') in the transcriptions. 
Our lexical features were extracted from these transcripts. 
We also collected a wide range of prosodic and facial features. 

\subsubsection{Prosodic Features}
\begin{table}
\footnotesize
  \caption{List of prosodic features and their brief descriptions}
  \renewcommand{\arraystretch}{1.2}
  \begin{tabular}{ ||  l  | l  ||}   \hline \hline
  Prosodic Feature & Description \\
  \hline \hline
	Energy       & Mean spectral energy.\\
	F0 MEAN 	& Mean F0 frequency. \\
	F0 MIN	& Minimum F0 frequency. \\
	F0 MAX 	& Maximum F0 frequency.\\
	F0 Range 	& Difference between F0 MAX and F0 MIN.\\
	F0 SD	& Standard deviation of F0.\\
	Intensity MEAN 	& Mean vocal intensity. \\
	Intensity MIN	& Minimum vocal intensity . \\
	Intensity MAX 	& Maximum vocal intensity . \\
	Intensity Range 	& Difference between max and \\
						& min intensity. \\
	Intensity SD	& Standard deviation.\\
	F1, F2, F3 MEAN & Mean frequencies of the first 3\\ 			 			   & formants: F1, F2, and F3.\\
	F1, F2, F3 SD	& Standard deviation of F1, F2, F3.  \\
	F1, F2, F3 BW 	& Average bandwidth of F1, F2, F3. \\
	F2/F1 MEAN & Mean ratio of F2 and F1.\\
	F3/F1 MEAN & Mean ratio of F3 and F1.\\
	F2/F1 SD & Standard deviation of  F2/F1.\\
	F3/F1 SD & Standard deviation of  F3/F1.\\
	Jitter 	& Irregularities in F0 frequency. \\
	Shimmer 	& Irregularities in intensity. \\
	Duration 	& Total interview duration. \\
	\% Unvoiced & Percentage of unvoiced region. \\
	\% Breaks		& Average percentage of breaks. \\
	maxDurPause	& Duration of the longest pause. \\
	avgDurPause	& Average pause duration.\\
	\hline
  \end{tabular}
\label{table:prosodic_features}
\end{table}
Prosody reflects our speaking style, particularly the rhythm and the intonation of speech. 
Prosodic features have been shown to be effective for social intent modeling~\cite{Soman-ICASSP10,Frick-PsyBultn85,Zechner09}. 
To distinguish between the speech of the interviewer and the interviewee, we manually annotated the beginning and end of each of the interviewee's answers. 
We extracted and analyzed prosodic features of the interviewee's speech. 
Each prosodic feature is first collected over an  interval corresponding to a single answer by the interviewee, and then averaged over all her/his five answers. We used the open-source speech analysis tool PRAAT~\cite{Boersma-Praat09} for prosody analysis.

The important prosodic features include pitch information, vocal intensities, characteristics of the first three formants, and spectral energy, which have been reported to reflect our social traits~\cite{Frick-PsyBultn85}. 
To reflect the vocal pitch, we extracted the mean and standard deviation of fundamental frequency F0 (F0 MEAN and F0 SD), the minimum and maximum values (F0 MIN, F0 MAX), and the total range (F0 MAX - F0 MIN). 
We extracted similar features for voice intensity and the first 3 formants. 
Additionally, we collected several other prosodic features such as pause duration, percentage of unvoiced frames, jitter (irregularities in pitch), shimmer (irregularities in vocal intensity), percentage of breaks in speech, etc. 
Table~\ref{table:prosodic_features} shows the complete list of prosodic features.

\subsubsection{Lexical features}
Lexical features can provide valuable information regarding the interview content and the interviewee's personality. 
One of the most commonly used lexical features is the unigram counts for each individual word. However, treating unigram counts as features often results in sparse high-dimensional feature vectors, and suffers from the ``curse of dimensionality'' problem, especially for a limited sized corpus. 
\begin{table}
\footnotesize
  \caption{LIWC Lexical features used in our system.}
  \renewcommand{\arraystretch}{1.2}
  \begin{tabular}{||  l  ||  l  ||}
  \hline \hline
  LIWC Category & Examples \\
  \hline \hline
	I 	&  \emph{I, I'm, I've, I'll, I'd,} etc.  \\
	We	& \emph{we, we'll, we're, us, our,} etc. \\
	They & \emph{they, they're, they'll, them,} etc.  \\
	Non-fluencies	&words introducing non-fluency in \\
				& speech, e.g., \emph{uh, umm, well}.  \\
	PosEmotion	& words expressing positive emotions, \\
					& e.g.,  \emph{hope,  improve, kind, love}. \\
	NegEmotion	& words expressing negative emotions, \\
					& e.g., \emph{bad, fool, hate, lose}. \\ 
	Anxiety 	& \emph{nervous, obsessed, panic, shy,} etc.  \\
	Anger	& \emph{agitate, bother, confront,  disgust,} etc.\\
	Sadness	& \emph{fail, grief, hurt, inferior,} etc.\\
	Cognitive	& \emph{cause, know, learn, make, notice,} etc. \\
	Inhibition	& \emph{refrain, prohibit, prevent, stop,} etc.\\
	Perceptual	& \emph{observe, experience, view, watch,} etc. \\
	Relativity	& \emph{first, huge, new,} etc.  \\
	Work	& \emph{project, study, thesis, university,} etc.\\
	Swear	& Informal and swear words.\\
	Articles 	& \emph{a, an, the,} etc.\\
	Verbs 	& common English verbs.\\
	Adverbs	& common English adverbs.\\
	Prepositions	& common prepositions. \\
	Conjunctions 	& common conjunctions. \\
	Negations 	& \emph{no, never, none, cannot, don't,} etc.\\
	Quantifiers & \emph{all, best, bunch, few, ton, unique,} etc.\\
	Numbers & words related to number, e.g., \\
			& \emph{first, second, hundred,} etc.\\
\hline			
  \end{tabular}
\label{table:liwc}
\end{table}

We address this challenge with two techniques. First, instead of using raw unigram counts, we employed counts of various psycholinguistic word categories defined by the tool ``Linguistic Inquiry Word Count'' (LIWC)~\cite{Pennebaker-liwc01}. 
The LIWC categories include words describing negative emotions (sad, angry, etc.), positive emotions (happy, kind, etc.), different function word categories (articles, quantifiers, pronouns, etc.), and various content categories (e.g., anxiety, insight). 
We selected 23 such LIWC word categories, which is significantly smaller than the number
of individual words.
The LIWC categories correlate with various psychological traits, and often provide indications about our personality and social skills~\cite{Tausczik-jlsp10}. 
Many of these categories are intuitively related to interview performance. Table~\ref{table:liwc} shows the complete list of the LIWC features used in our experiments.

Although the hand coded LIWC lexicon has proven to be useful for modeling many different social behaviors, the lexicon is predefined and may not cover many important aspects of job interviews. 
To address this challenge, we aimed to automatically learn a lexicon from the interview dataset. 
We apply the  Latent Dirichlet Allocation (LDA)~\cite{Blei-jmlr03} method to automatically learn common topics from our interview dataset. 
We set the number of topics to 20. 
For each interview, we estimate the relative weights of these learned topics, and use these weights as lexical features. 
Similar ideas have been exploited by Ranganath et al.~\cite{Ranganath-emnlp09,Ranganath-CSL13} for modeling social traits in speed dating dataset, but they used deep auto-encoders~\cite{Hinton-AAAS06} instead of LDA. 

Finally, we collected additional lexical features that correlate to job interview ratings. 
These are features related to our linguistic and speaking skills. Table~\ref{table:wpsec} contains the full list. 
Similar speaking rate and fluency features were exploited by Zechner et al.~\cite{Zechner09} in the context of automated scoring of non-native speech in TOEFL practice tests.
\begin{table}
  \caption{Additional features related to speaking rate and fluency.}
    \renewcommand{\arraystretch}{1.2}
    \centering
  \begin{tabular} { || l || l ||} \hline \hline
  	feature Name	& Description \\ \hline \hline
	wpsec  	& Words per second.  \\
	upsec 	& Unique words per second.\\
	fpsec	& Filler words per second.\\
	wc		& Total number of words. \\
	uc		& Total number of unique words. \\ \hline
  \end{tabular}
\label{table:wpsec}
\end{table}

\subsubsection{Facial features}
We extracted facial features for the interviewees from each frame in the video. 
First, faces were detected using the Shore~\cite{Froba-FG04} framework. 
We trained a classifier to distinguish between neutral and smiling faces. The classifier is trained using the AdaBoost algorithm. 
The classifier output is normalized in the range [0,100], where 0 represents no smile, and 100 represents full smile. 
Finally, we averaged the smile intensities from individual frames, and used this as a feature in our model. 
We also extracted head gestures such as nods and shakes as explained in~\cite{Hoque-ubicomp13}. 

\begin{figure}
\centering
  \includegraphics[width=0.35\textwidth]{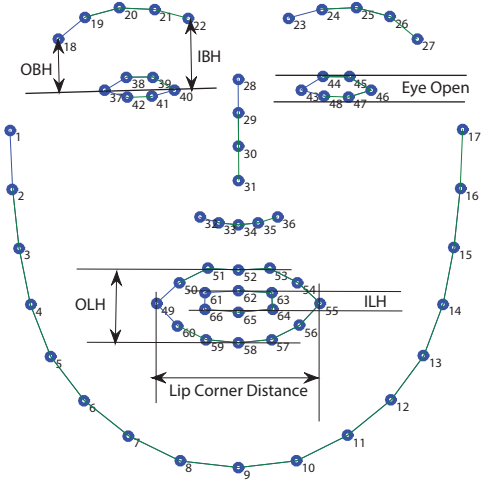}
\caption{Illustration of facial features: OBH (outer eye-brow height), IBH (inner eye-brow height), OLH (outer lip height), ILH (inner lip height), eye opening, and LipCDT (lip corner distance).}
\label{fig:face_annot}
\end{figure}

In addition to the smile intensity and head gestures (nod and shake), we also extracted a number of other facial features using a Constrained Local Model (CLM)~\cite{saragih2009face} based face tracker\footnote{https://github.com/kylemcdonald/FaceTracker}, as illustrated in Fig~\ref{fig:face_annot}. 
The face tracker detects 66 interest points on a face image. 
It works by fitting the following parametric shape model~\cite{saragih2009face}\cite{Saragih2011}:
\begin{equation}
\mathbf{x}_i = s\mathbf{R}(\mathbf{\bar{x}}_i + \mathbf{\Psi_i}\mathbf{q})+\mathbf{t}, 
\end{equation}
where $\mathbf{x}_i$ is the coordinate of $i^{\text{th}}$ interest point and $\mathbf{\bar{x}}_i$ denotes its mean location pre-trained from a large collection of hand-labeled training images. 
$\mathbf{\Psi_i}$ denotes the bases of local variations for the $i^{\text{th}}$ interest point. 
Each element of the vector $\mathbf{q}$ represents a coefficient corresponding to a basis of local variation. 
The parameters $s, \mathbf{R},$ and $t$ corresponds to the global transformations associated with scaling, rotation, and translation respectively. 
The face tracker adjusts the model parameters $p=\{s,\mathbf{R},\mathbf{q},\mathbf{t}\}$ so that each of the mean interest points ($\mathbf{\bar{x}}_i$) fits best to its corresponding point ($\mathbf{x}_i$) on the test face. 

While extracting features from these tracked interest points, we want to disregard the global transformations (translation, rotation, and scaling), and consider only the local transformations, which provide useful information regarding our facial expressions. 
After the face tracker converges to an optimal estimate of the parameters, we recalculate each of the interest points $\mathbf{x}_i$ by applying the local transformations only, while disregarding the global transformations ($s, \mathbf{R},$ and $t$).
Mathematically, we calculate the following shape model from the optimal parameters obtained from the face tracker:
\begin{equation}
\mathbf{\hat{x}}_i = (\mathbf{\bar{x}}_i + \mathbf{\Psi_i}\mathbf{q}) 
\end{equation}
Once we find $\mathbf{\hat{x}}_i$, we calculate the distances between the corresponding interest points to find out the features OBH (outer eye-brow height), IBH (inner eye-brow height), OLH (outer lip height), and ILH (inner lip height), eye opening, and LipCDT (lip corner distance), as illustrated in Figure~\ref{fig:face_annot}. 
By disregarding the global transformation parameters, the extracted facial features are invariant to global translations, rotations, and scaling variations. 
In addition to the features shown in Figure~\ref{fig:face_annot}, we separately incorporated three head pose features (Pitch, Yaw and Roll), based on the corresponding elements of the rotation matrix $\mathbf{R}$.

\subsubsection{Feature Normalization}
We concatenate the three types of features described above and obtain one combined feature vector. To remove any possible bias related to the range of values associated with a feature, we normalized each feature to have zero mean and unit variance, which allows treating all the features uniformly.

\subsection{Ground Truth Ratings and Turker Quality Estimation}
\label{section:turkerquality}
We aim to automatically estimate the reliability of each Turker, and the ground truth ratings based on the Turkers' ratings. We adapt a simplified version of the existing latent variable models~\cite{raykar2010learning} that treat each Turker's reliability and ground truth ratings as latent parameters, estimate their values using an EM-style iterative optimization technique.

Let us assume an input training dataset $\mathcal{D} = \{\mathbf{x}_i, y_i\}_{i=1}^N$ 
containing $N$ feature vectors $\mathbf{x}_i$ (one for each interview video), for which the ground truth label $y_i$ is not known. 
Instead we acquire subjective labels $\{y_i^1, \ldots, y_i^K \}$ from $K$ Turkers on seven point Likert scale, i.e., $y_i^j \in \{1,2,\ldots, 7\}$. 
Given this dataset $\mathcal{D}$, our goal is to learn the true rating ($y_i$) and the reliability of each worker ($\lambda_j$).

To simplify the estimation problem, we assume the Turkers' ratings to be real numbers, i.e., $y_i^j \in \mathbb{R}$. We also assume that each Turker's rating is a noisy version of the true rating $y_i \in \mathbb{R}$, perturbed via additive Gaussian noise. Therefore, the probability distribution for the $y_i^j$:
\begin{equation}
Pr[y_i^j | y_i, \lambda_j] = \mathcal{N} (y_i^j | y_i, 1/\lambda_j)
\end{equation}
where $\lambda_j$ is the unknown inverse-variance and the measure of reliability for the $j^{th}$ Turker. By taking logarithm on both side and ignoring constant terms, we get the log-likelihood function:\
\begin{equation}
L = \sum_{i=1}^N \sum_{j=1}^K \left[ \frac{1}{2} \log{\lambda_j} - \frac{\lambda_j}{2} (y_i^j - y_i)^2 \right]
\end{equation}

The log-likelihood function is non-convex in $y_i$ and $\lambda_j$ variables. However, if we fix $y_i$, the log-likelihood function becomes convex with respect to $\lambda_j$, and vice-versa. Assuming $\lambda_j$ fixed, and setting $\frac{\partial L}{\partial y_i} = 0$, we obtain the update rule:
\begin{equation}
y_i = \frac{ \sum_{j=1}^K  \lambda_j y_i^j} { \sum_{j=1}^K  \lambda_j }
\end{equation}
Similarly, assuming $y_i$ fixed, and setting $\frac{\partial L}{\partial \lambda_j} = 0$, we obtain the update rule:
\begin{equation}
\lambda_j = \frac{ \sum_{i=1}^N  (y_i^j - y_i)^2 } {N}
\end{equation}

We alternately apply the two update rules for $y_i$ and $\lambda_j$ for $i = 1, \ldots, N$ and $j = 1, \ldots, K$ until convergence. After convergence, the estimated $y_i$ values are treated as ground truth ratings and used for training our prediction models.

\subsection{Score Prediction from Extracted Features}
Using the features described in the previous section, we train regression models to predict the interview scores. We also train models to predict other interview-specific traits such as excitement, friendliness, engagement, awkwardness, etc. We experimented with many different regression models: Support Vector Machine Regression (SVR)~\cite{Smola-04}, Lasso~\cite{Tibshirani-Lasso96}, L$_1$ Regularized Logistic Regression, Gaussian Process Regression, etc. We will only discuss SVR and Lasso, which achieved the best results with our dataset.

\subsubsection{Support Vector Regression (SVR)}
The Support Vector Machine (SVM) is a widely used supervised learning method. In this paper, we focus on the SVMs for regression, in order to predict the performance ratings from interview features. Suppose we are given a training data $\{  (\mathbf{x}_1, y_1),  \ldots ,(\mathbf{x}_N, y_N))\}$, where $\mathbf{x}_i \in \mathbb{R}^d$ is a $d$-dimensional feature vector for the $i^{th}$ interview in the training set. For each feature vector $\mathbf{x}_i$, we have an associated value $y_i \in \mathbb{R}_+$ denoting the interview rating. Our goal is to learn the optimal weight vector $\mathbf{w} \in \mathbb{R}^d$ and a scalar bias term $b \in \mathbb{R}$ such that the predicted value for the feature vector $\mathbf{x}$ is: $\hat{y} = \mathbf{w}^T \mathbf{x} + b$.  We minimize the following objective function:
\begin{equation}
\begin{aligned}
& \underset{\mathbf{w}, \xi_i, \hat{\xi}_i, b}{\text{minimize}}
& & \frac{1}{2}  \|  \mathbf{w} \|^2  + C \sum_{i = 1}^N (\xi_i + \hat{\xi}_i)\\
& \text{subject to}
& & y_i - \mathbf{w}^T \mathbf{x}_i - b \leq \epsilon + \xi_i,  \ \forall i \\
&&&  \mathbf{w}^T \mathbf{x}_i + b - y_i \leq \epsilon + \hat{\xi}_i,  \ \forall i \\
&&&  \xi_i, \hat{\xi}_i \geq 0,  \ \forall i \\
\end{aligned}
\end{equation}
The $\epsilon \geq 0$ is the precision parameter specifying the amount of deviation from the true value that is allowed, and $(\xi_i, \hat{\xi}_i)$ are the slack variables to allow deviations larger than $\epsilon$. The tunable parameter $C  > 0$ controls the tradeoff between goodness of fit and generalization to new data. The convex optimization problem is often solved by maximizing the corresponding dual problem. In order to analyze the relative weights of different features, we transform it back to the primal problem and obtain the optimal weight vector $\mathbf{w}^*$ and bias term $b^*$. The relative importance of the $j^{th}$ feature can be interpreted by the associated weight magnitude $|w_j^*|$.

\subsubsection{Lasso}
The Lasso regression method aims to minimize the residual prediction error in the presence of an $L_1$ regularization function. Using the same notation as the previous section, let the training data be $\{  (\mathbf{x}_1, y_1),  \ldots ,(\mathbf{x}_N, y_N))\}$. Let our linear predictor be of the form: $\hat{y} = \mathbf{w}^T \mathbf{x} + b$. The Lasso method estimates the optimal $\mathbf{w}$ and $b$ by minimizing the following objective function:
\begin{equation}
\begin{aligned}
& \underset{\mathbf{w},b}{\text{minimize}}
& &\sum_{i=1}^N \left( y_i -  \mathbf{w}^T \mathbf{x}_i - b \right)^2 \\
& \text{subject to}
& & \| \mathbf{w} \|_1 \leq \lambda \\
\end{aligned}
\end{equation}
where $\lambda > 0$ is the regularization constant, and  $\| \mathbf{w} \|_1 = \sum_{j=1}^d |w_j|$ is the $L_1$ norm of $\mathbf{w}$. The $L_1$ regularization is known to push the coefficients of the irrelevant features down to zero, thus reducing the predictor variance. We control the amount of sparsity in the weight 
vector $\mathbf{w}$ by tuning the regularization constant $\lambda$.
\section{Results}
\label{section:results}
We organize our results in two sections. First, we analyze the ratings by Mechanical Turk workers (Section~\ref{section:turkresults}). The quality and reliability of Turkers' ratings are assessed by observing how well the Turkers agree with each other (Section~\ref{section:interrater}).  
In addition, we identify which traits are important to succeed in job interviews by measuring the correlations of the ratings for individual traits with the overall ratings (Section~\ref{section:correlation:traits}). 
Furthermore, we examine the correlations between the ratings for individual video segments with that for the entire videos.  This allowed us to evaluate the temporal patterns in job interviews (Section~\ref{section:correlation:temporal}).

In Section~\ref{section:prediction}, we present the prediction accuracies for the trained regression models (SVR and Lasso) based on automatically extracted features, and analyze the relative influence of different modalities and features on prediction accuracy. 
%
\subsection{Analysis of Mechanical Turk Dataset}
\label{section:turkresults}
\subsubsection{Inter-Rater Agreement}
\label{section:interrater}
\begin{figure}
\centering
\epsfig{file=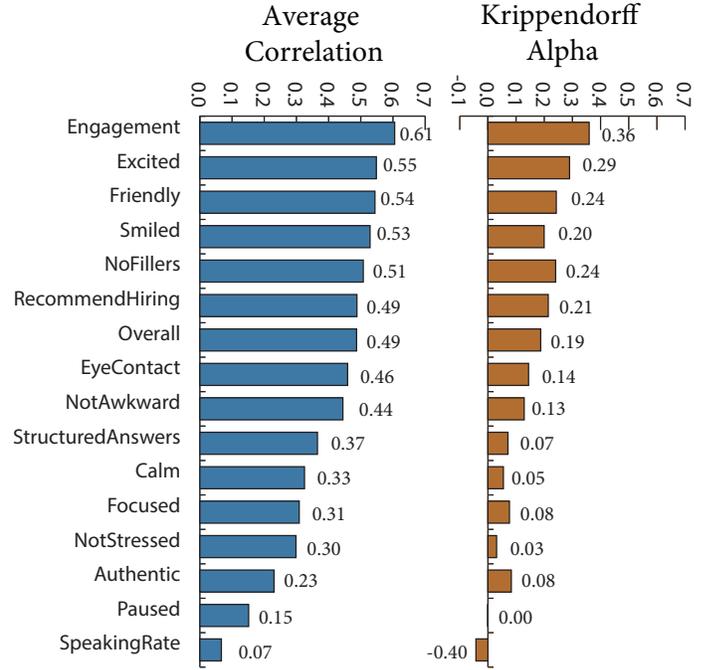, width = 0.5\textwidth}
\caption{The inter-rater agreement among the turkers, measured by the Krippendorff's Alpha (varies in the range $[-1,1]$) and the average one-vs-rest correlation of their ratings (range $[-1,1]$).}\label{fig:InterRaterAgreement}
\end{figure}
To assess the quality of the ratings, we calculate Krippendorff's Alpha \cite{Krippendorff1970} for each trait. 
In this case, Krippendorff's Alpha is more meaningful than the frequently used Fleiss' Kappa~\cite{Fleiss98}, as the ratings are ordinal values (on a 7-point Likert scale). 
The value of Krippendorff's Alpha can be any real number in the range $[-1,1]$, with 1 being the perfect agreement and -1 being absolute disagreement among the raters. 
We also estimate the correlation of each Turker's rating with the mean rating by the other Turkers for each trait. 
Figure \ref{fig:InterRaterAgreement} shows that some traits have relatively good inter-rater agreement among the Turkers (e.g., ``engagement'', ``excitement'', ``friendliness''). Some other traits such as: ``stress", ``authenticity'', ``speaking rate'', and ``pauses'' have low inter-rater agreement. This may be because the Turkers were not in a position to judge those categories with the video data only.

\subsubsection{Correlation among the Behavioral Traits}
\label{section:correlation:traits}
\begin{figure}
\centering
\epsfig{file=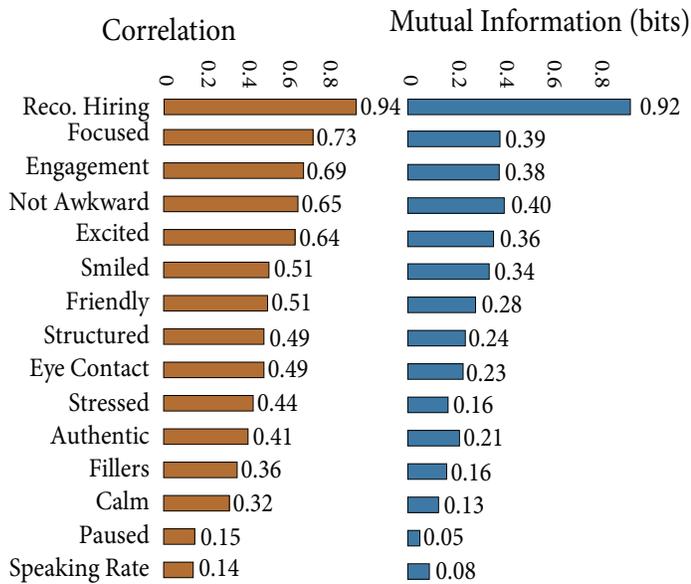, angle = -90, width = 0.5\textwidth}
\caption{Correlation and Mutual information between overall rating and ratings on other traits.}
\label{fig:relativeImportanceOfTraits}
\end{figure}

We are interested in identifying the traits that correlate highly with overall ratings. 
This knowledge can help interviewees understand the most important behavioral traits in job interviews. 
We plot the mutual information and correlation between various ratings given by the Mechanical Turk workers and the overall rating of the interviewee performance in Figure~\ref{fig:relativeImportanceOfTraits}.  

The first bar in Figure~\ref{fig:relativeImportanceOfTraits} represents whether the rater will recommend hiring the interviewee. 
It is another form of the overall rating and shows high correlation and mutual information with the overall rating. 
It is evident from the plot that the most important trait in an interview is to stay focused. 
This trait shows a 73\% correlation with the overall rating. 
Some other top traits include possessing an engaging tone, not appearing awkward, being excited, and displaying an appropriate smile. 
The mutual information and correlation coefficient closely follow the patterns. 
This plot gives us an insight into what constitutes a good interview.   

\subsubsection{First (and Last) Impression Matters}
\label{section:correlation:temporal}
\begin{figure*}
	\centering
	\subfigure[Correlation and Mutual Information for overall performance]{
		\includegraphics[width=0.3\textwidth]{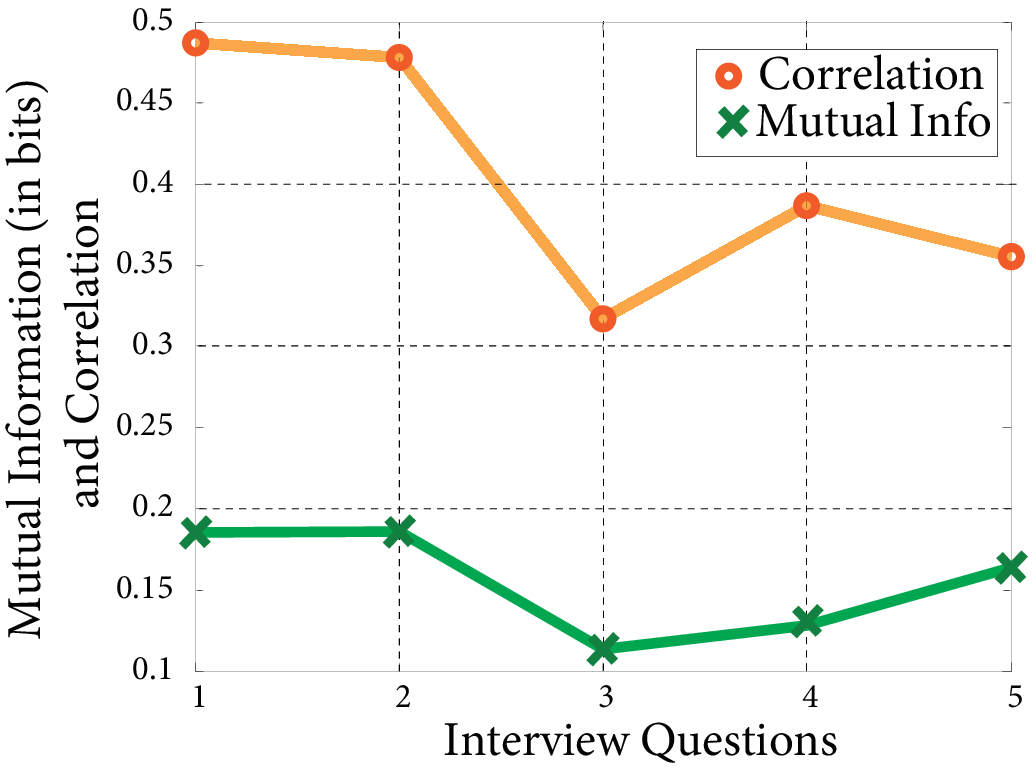}
		\label{fig:firstImpressionMatters_overall}
	}
	\subfigure[Traits following patterns similar to the overall performance]{
		\includegraphics[width=0.29\textwidth]{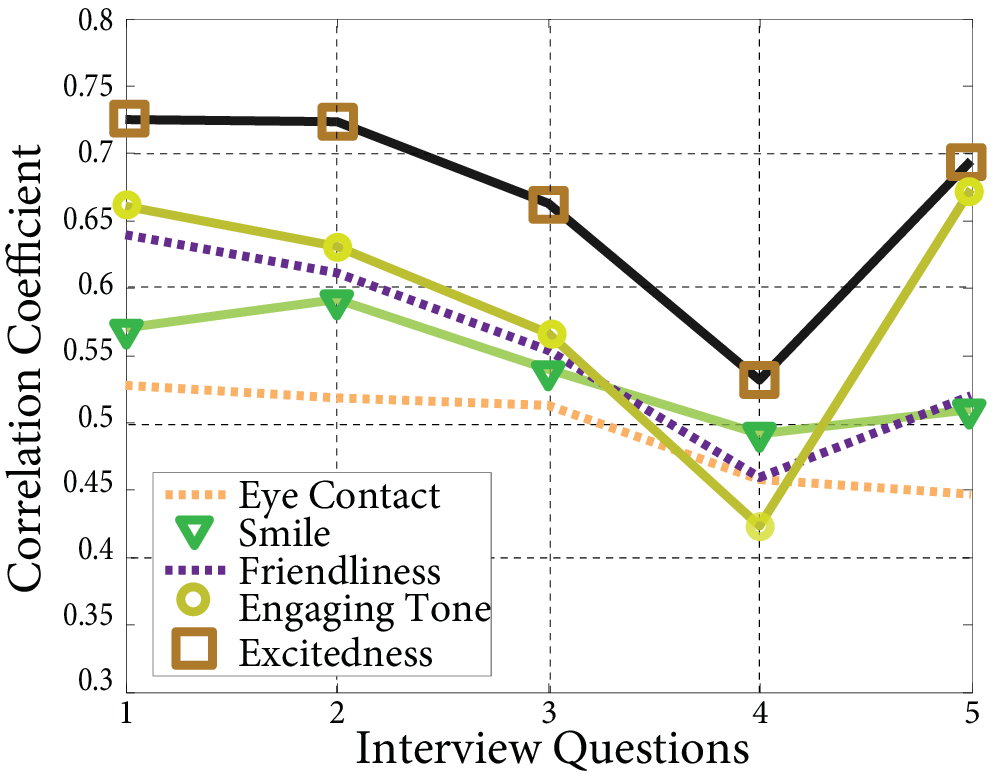}
		\label{fig:firstImpressionMatters_good}
	}
	\subfigure[Traits not following patterns]{
		\includegraphics[width=0.32\textwidth]{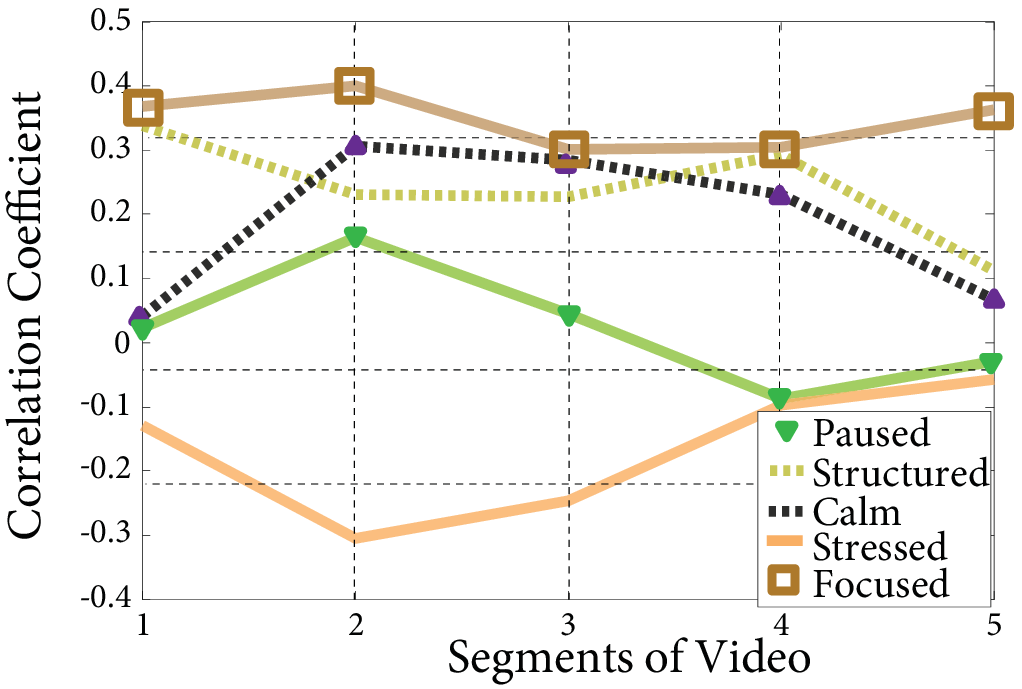}
		\label{fig:firstImpressionMatters_bad}		
	}
	\caption{Correlation between ratings of different segments and the rating on the whole interview.}
	\label{fig:first_impression}
\end{figure*}

We would like to understand how the performance in different interview questions during an interview affects the overall rating. 
To understand this temporal relationship, we calculated the correlation and mutual information between the ratings for each individual interview question and the ratings for the entire videos. 
In Figure \ref{fig:first_impression}, we plot this relationship. 
It is evident from Figure~\ref{fig:firstImpressionMatters_overall} that performance on the first question correlates most with the overall performance. 
After the first question, the correlation gradually decays. 
We can interpret this result as follows: If an interviewee performs well for the first question, it is more likely that he/she will end up receiving an above average rating. 
It is true in the opposite case as well; if an interviewee performs poorly in the first question, he/she is more likely to receive a poor overall rating. This finding is also supported by existing evidence from psychological point of view~\cite{Dougherty94,Curhan07}. 

A similar pattern of \emph{first impression matters} holds for ratings on various other traits of the interviewee's behavior, such as whether he/she was excited, smiled, maintained eye contact, talked in engaging tone, or even appeared friendly. 
Figure \ref{fig:firstImpressionMatters_good} illustrates this. 
We notice from this figure that there is a sudden spike in correlation for the last question. 
This indicates the fact that, although the first question matters the most, the interviewee can significantly change the interviewer's perception during the response to the final question.

Figure \ref{fig:firstImpressionMatters_bad} shows some traits (e.g., pause, calmness, stress) do not follow the pattern discussed above. 
However, they have very low correlation values to begin with. 
We believe it is difficult for Mechanical Turk workers to accurately judge these traits as these judgments demand considerable concentration.

We need to be cautious while interpreting this result.
Although the ratings for the first question had maximum correlation with the overall ratings for the entire interview, we can not say whether it is due to the temporal order or the verbal content of the question itself. 
However, we would like to emphasize that our mock interviews start with a question about interviewee's background, which is consistent with many real-world job interviews. 
\subsection{Prediction using Automated Features}
\label{section:prediction}

\subsubsection{Prediction Accuracy using Trained Models}
\label{section:accuracy}

Given the feature vectors associated with each interview video, we would like to provide feedback to users about their overall performance in the interview, the likelihood of getting an offer, and insights into other personality traits that are relevant for job interviews. We train regression models for predicting ratings for a total of 16 traits or rating categories (as shown in Table~\ref{table:traits}). 

The entire dataset has a total of 138 interview videos (for the 69 participants, 2 interviews for each participant). We used 80\% of the videos for training, and the remaining 20\% for testing. To avoid any artifacts related to how we split the data, we performed 1000 random trials. In each trial, we randomly select 80\% videos for training, and use the rest for testing. 
We report our results averaged over these 1000 independent trials. 
In each trial, we trained 16 different regression models for all 16 traits. 
For each of the traits, we used exactly the same set of features. 
The model automatically learned the weights for individual features for each trait.

\begin{figure}[h]
\centering
  \includegraphics[width=0.52\textwidth]{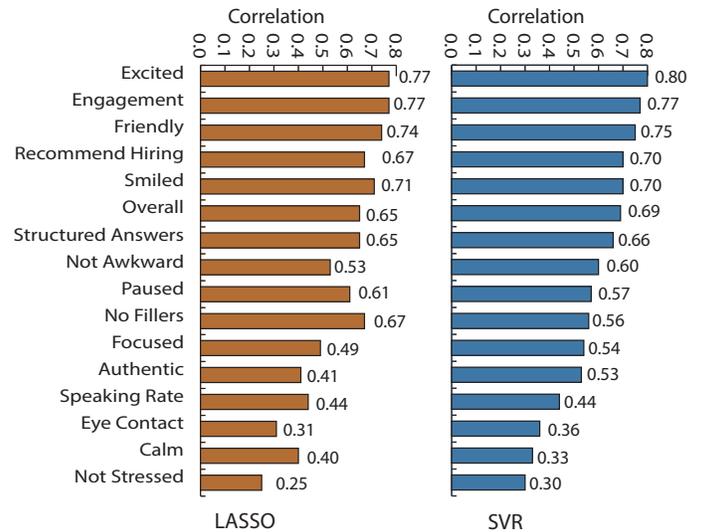}
\caption{Regression coefficients using two different methods: Support Vector Machine (SVM) and Lasso.}
\label{fig:results}
\end{figure}

We measure prediction accuracy by the correlation coefficients between the true ratings and predicted ratings in the test set. 
Figure~\ref{fig:results} displays the correlation coefficients for different traits, both with SVM and Lasso. 
The traits are shown in the order of their correlation coefficients. 
We observe that we can predict several traits with 0.75 or higher correlation coefficients: engagement, excitement, and friendliness. 
Furthermore, we performed well in predicting overall performance and hiring recommendation scores ($r ~ 0.70$ for SVM), which are the two most important scores for  interview decision.

We also evaluate the learned regression models for a two-class classification task. 
For each trait, we split the interviews into two groups by the median value for that trait. 
Any interview with a score higher than the median value for a particular trait is considered to be in the positive class (for that trait), and the rest are placed in the negative class. 
We then vary the threshold on the predicted scores by our regression models in the range $[1,7]$, and estimate the area under the Receiver Operator Curve (ROC). 
The baseline area under the curve (AUC) value is 0.50, as we split the classes by the median value. 
The AUC values for the learned models are presented in Table~\ref{table:auc}. 
Again, we observe high accuracies for engagement, excitement, friendliness, hiring recommendation, and the overall score ($AUC > 0.80$ for SVM).
\begin{table}
\centering
\caption{The average area under the ROC curve.}
\label{table:auc}
\begin{tabular}{|| l || l || l ||}  \hline \hline
%
Trait & SVM & Lasso \\ \hline \hline
Excited& 0.904 & 0.885 \\ 
Engagement & 0.858 & 	0.850 \\
Smiled	& 0.845	& 0.845\\
Friendly	& 0.824 &	0.793\\
Recommend Hiring &	0.815 & 0.796 \\
Structured Answers & 0.812 & 0.799 \\
Not Awkward & 0.808 &	0.787 \\
Overall	& 0.805 & 0.777\\
No Fillers &	0.803 & 0.855\\
Focused &	0.791 & 0.677 \\
Paused &	0.749 & 0.749 \\
Authentic	& 0.688 &	0.642\\
Eye Contact & 0.676 & 0.622\\
Calm 	& 0.651 &	0.669\\
Speaking Rate &	0.608 & 0.546 \\
Not Stressed & 0.604 &	0.572\\ \hline
\end{tabular}
\end{table}

When we examine the traits with lower prediction accuracy, we observe: (1) either we have low interrater agreement for these traits, which indicates unreliable ground truth data  (e.g., calm, stressed, structured answer, pause, etc.), or (2) we lack key features necessary to predict these traits (e.g., eye contact). In the absence of eye tracking information (which is very difficult to obtain automatically), we do not have enough informative features to predict eye contact.

\subsubsection{Feature Analysis} 
\label{section:featureanalysis}
\begin{figure*}
\centering
\subfigure[Relative proportion of the top twenty prosodic, lexical and facial (smile) features as learned by SVM and LASSO classifiers. The weights semantically match our perceptions on the traits]{
\includegraphics[width=0.80\textwidth]{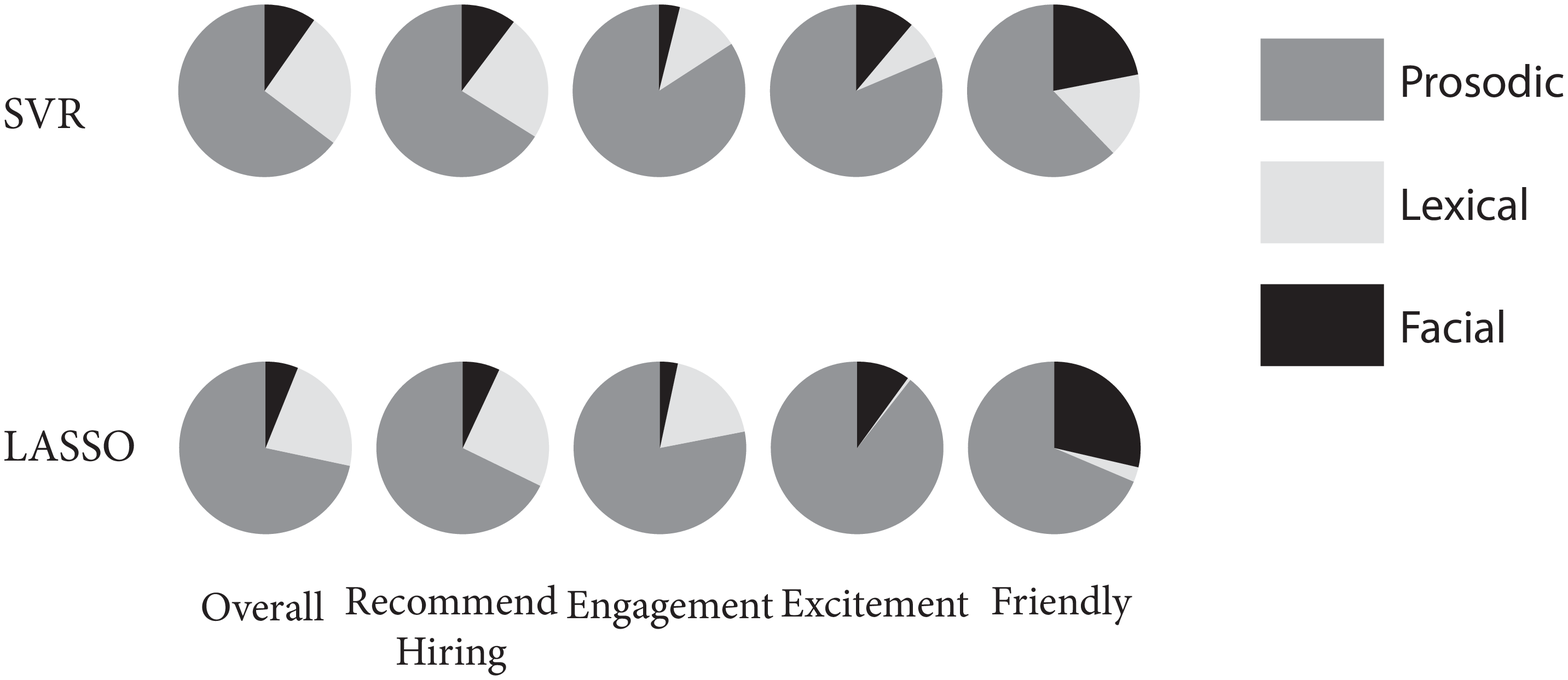}
		\label{fig:relWeights}}
\subfigure[Correlation Coefficients for SVM, for different combinations of facial (F), prosodic (P), and lexical (L) features]{
\includegraphics[width=0.95\textwidth]{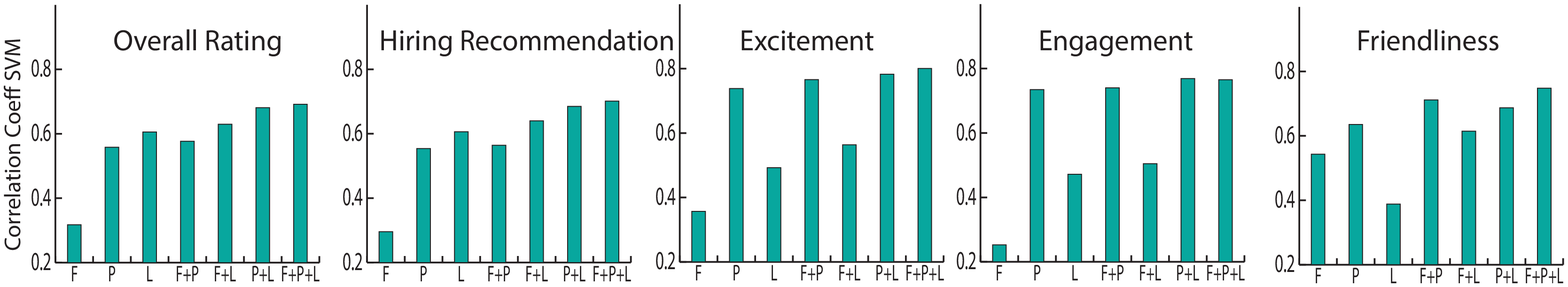}
		\label{fig:featurecomb}}
\caption{Analysis of relative importance of facial, prosodic, and lexical features. }
\end{figure*}

The relative weights of individual features in our regression model can provide valuable insights on essential constituents of a job interview. 
To analyze this, we observed the features with highest weights for the SVM and the Lasso model. 
We considered five traits with high accuracy: overall score, recommend hiring, excitement, engagement, and friendliness. 
We considered the top twenty features in the order of descending weight magnitude, and estimate the summation of the weight magnitudes of the features in each of the three categories: prosodic, lexical, and facial features. 
The relative proportion of prosodic, lexical and facial features are illustrated in Figure~\ref{fig:relWeights}, which shows that both SVM and Lasso assign higher weights to prosodic features while predicting engagement and excitement. 
This indicates that engagement and excitement are expressed through prosodic features, which agrees with our intuition. 
For both models, the relative weights of features for predicting the ``overall rating" and ``recommend hiring" are similar, which is expected, as these two traits are highly correlated (Figure~\ref{fig:relativeImportanceOfTraits}). 
Since we had smaller number of facial features, the relative weights for facial features is much lower. 
However, facial features, particularly the smile, were found significant for predicting friendliness. This result provides a solid ground for claiming that smile is very important in order to appear friendly.
Figure~\ref{fig:featurecomb} shows the importance of using multimodal features for predicting social traits in job interviews. 
In most cases, the best correlation coefficient was obtained when we incorporated all three modalities. 
Although lexical features were critical for predicting overall ratings and likelihood of getting hired, they were not strong predictors of excitement, engagement, and friendliness. Prosodic features played important role for predicting all the five traits, indicating that our speaking style plays a critical role in job interviews.  

\subsubsection{Recommendation from our Framework}
\label{section:recommendation}
To better understand the recommended behavior in job interviews, we analyze the feature weights in our regression model. The weights with positive signs and higher magnitudes can potentially indicate elements of a successful job interview. 
The negative weights, on the other hand, indicates behaviors we should avoid. 

We sort the features by the magnitude of their weights and list the top twenty features (excluding the topic features) in Table~\ref{table:features:svm}. 
We see from this table that people having higher speaking rate (higher words per second (\emph{wpsec}), total number of words (\emph{wc}), and total number of unique words (\emph{uc}), etc.) are perceived as better candidates in a job interview. People who speak more fluently and use less filler words (lower number of filler words per second (\emph{fpsec}), total number of filler words (\emph{Fillers}), total number non-fluency words (\emph{Non-fluencies}), less unvoiced region in speech (\emph{\%Unvoiced}), and fewer breaks in speech (\emph{\%Breaks})) are perceived as better candidates. We also find that higher interview score correlates with higher usage of words in LIWC category \emph{They} (e.g. they, they'll, them, etc.) and lower usage of words related to \emph{I}. 
The overall interview performance and likelihood of hiring correlate positively with proportion of positive words, and negatively with proportions of negative words, which agrees with our experience. 
Individuals who smiled more performed better in job interviews. Finally, those speaking with a higher proportion of quantifiers (e.g., best, every, all, few), perceptual words (e.g. see, observe, know), and other functional word classes (articles, prepositions, conjunctions) obtained higher scores in interview. 
As we saw earlier, features related to prosody and speaking style are more important to appear excited and engaged. 
Particularly the amplitude, variations in the voice intensity, and the first 3 formants had high positive weights in our prediction model. Finally, besides smiling, people who spoke more words related to \emph{``We''} than \emph{``I''} were perceived as friendlier.

\begin{table*}
\caption{Feature Analysis using the SVM model. We are listing the top twenty features ordered by their weight magnitude. We have  excluded the topic features for the ease of interpretation.}
\label{table:features:svm}
\footnotesize 
\begin{tabular}{|l  l |l  l |l  l |l  l |l  l |}
\hline
Overall &  & RecommendHiring &  & Excited &  & EngagingTone &  & Friendly& \\ \hline 
avgBand1 & -0.111 & wpsec & 0.138 & avgBand1 & -0.152 & intensityMax & 0.175 & smile & 0.239\\ 
wpsec & 0.104 & avgBand1 & -0.134 & diffIntMaxMin & 0.134 & avgBand1 & -0.168 & mean pitch & 0.156\\ 
Fillers & -0.085 & Fillers & -0.126 & wpsec & 0.13 & diffIntMaxMin & 0.155 & f3STD & -0.11\\ 
Quantifiers & 0.084 & percentUnvoiced & -0.116 & intensityMax & 0.125 & intensityMean & 0.144 & LipCDT & 0.1\\ 
avgDurPause & -0.081 & smile & 0.101 & nod & 0.121 & wpsec & 0.132 & intensityMax & 0.098\\ 
smile & 0.079 & PercentBreaks & -0.094 & mean pitch & 0.118 & avgBand2 & -0.112 & diffIntMaxMin & 0.095\\ 
upsec & 0.078 & upsec & 0.088 & smile & 0.117 & f1STD & -0.107 & intensityMean & 0.087\\ 
percentUnvoiced & -0.076 & avgDurPause & -0.088 & f3STD & -0.11 & f2STDf1 & 0.101 & f1STD & -0.086\\ 
f3meanf1 & 0.075 & intensityMean & 0.085 & intensityMean & 0.11 & Quantifiers & 0.092 & wpsec & 0.085\\ 
Relativity & 0.074 & nod & 0.085 & f1STD & -0.108 & intensitySD & 0.092 & Adverbs & 0.081\\ 
Positive emotion & -0.073 & f1STD & -0.08 & percentUnvoiced & -0.107 & f3meanf1 & 0.091 & I & -0.08\\ 
nod & 0.069 & Prepositions & 0.078 & PercentBreaks & -0.099 & f3STD & -0.085 & shimmer & -0.077\\ 
PercentBreaks & -0.067 & Positive emotion & -0.077 & intensitySD & 0.092 & smile & 0.085 & fmean3 & 0.077\\ 
maxDurPause & -0.066 & f3meanf1 & 0.077 & f2STDf1 & 0.092 & Cognitive & 0.083 & percentUnvoiced & -0.073\\ 
f1STD & -0.065 & Quantifiers & 0.075 & wc & 0.089 & upsec & 0.083 & PercentBreaks & -0.071\\ 
Prepositions & 0.063 & wc & 0.074 & Adverbs & 0.081 & percentUnvoiced & -0.079 & max pitch & 0.071\\ 
intensityMean & 0.061 & max pitch & 0.07 & f3meanf1 & 0.081 & PercentBreaks & -0.075 & avgBand1 & -0.07\\ 
f2STDf1 & 0.06 & uc & 0.07 & Cognitive & 0.08 & max pitch & 0.074 & nod & 0.07\\ 
uc & 0.059 & Articles & 0.069 & f2meanf1 & 0.078 & f2meanf1 & 0.07 & Sadness & 0.069\\ 
f2meanf1 & 0.058 & maxDurPause & -0.069 & avgBand2 & -0.074 & Adverbs & 0.069 & Cognitive & 0.064\\ 
\hline\end{tabular}
\end{table*}

\section{Discussion and Conclusion}
\label{section:discussions}
We present an automated prediction framework for quantifying social skills for job interviews. 
The proposed model shows encouraging results and predicts human interview ratings with correlation $r > 0.65$ and AUC $\sim 0.80$ (compared to the baseline AUC $= 0.50$). 
Several traits such as engagement, excitement, and friendliness were predicted with even higher accuracy ($r \sim 0.75$, AUC $> 0.85$). 
One of our immediate next steps will be to integrate the proposed prediction module with existing automated conversational systems such as MACH to allow valuable real-time feedback to the users.

To our knowledge, the interview dataset used in our experiments is the largest collection of job interview videos, collected under reasonably realistic settings. 
The interviews are conducted by professional career counselors.
We included the questions that would be relevant in most real-world job interviews. 
Despite efforts to record interviews in realistic settings, we do need to acknowledge several caveats and trade-offs.

All the participants in our dataset were MIT undergraduates, all of junior status, which may introduce a selection bias in our data. 
In future, we plan to conduct a more comprehensive study over a more general and diverse population group.
We deliberately chose not to specify a job description to encourage larger number of student participants.
At the time of the study, there were nearly 1000 junior students present at MIT, and nearly 30\% were international students. 
Out of the remaining 700 native English speaking juniors, we were able to recruit 90, which would have been difficult if we had limited our study to a specific job description. 
However, in the absence of a specific job description, the ground truth ratings may not necessarily correspond to actual hiring decisions, and may show a stronger bias towards non-verbal cues, as there is no specific skill requirements.
Furthermore, our mock interviews may lack the stress present in a real job interviews.
Although we promised to forward the resumes of the top 5\% candidates to several sponsor organizations, the incentive was not as strong as an actual job offer.
In the future, we would like to conduct  more controlled experiments with a specific job description and with stronger incentives to induce stress and competition.

We aimed to rate each video with multiple independent judges to avoid personal bias. 
As a first step, we recruited Turkers as this was scalable, quick, and less expensive. 
To ensure reliable ground truth ratings, each video was rated using 9 Mechanical Turk workers, and aggregated using the EM algorithm taking the reliability of each worker into account.
However, Turkers' ratings may not correspond to professional experts. 
In future, we plan to collect ratings from a panel of experts, and re-validate the results.

Interestingly, while training regression models using SVR, we obtained better prediction accuracy using the linear kernel, compared to other non-linear kernels (e.g., quadratic, cubic, or Gaussian kernels). 
This may indicate that our features do not exhibit complicated non-linear interactions.
However, the features used in the current models were mostly aggregated features, averaged over the entire duration of the video (e.g., average pitch, average smile intensity). 
It is plausible that our smile and intonation while uttering a specific word can be a determinant of the final interview decision.
The current aggregated features are incapable of modeling such temporal interactions. 
Modeling fine-grained temporal features across multiple modalities is left as our future work.

The outcome of job interviews often depends on a subtle understanding of the interviewee's response. 
In our dataset, we noticed interviews in which a momentary mistake (e.g., the use of a swear word) ruined the interview outcome. 
Due to the rare occurrences of such events, it is difficult to model these phenomena, and perhaps anomaly detection techniques could be more effective instead. 
Extending our prediction framework for quantifying these diverse and complex cues in job interviews can provide valuable insight and understanding regarding job interviews and human behavior in general.

Caveats aside, the results presented in this article show the importance of including multiple modalities while analyzing our social interactions. 
The analysis of the feature weights learned by our prediction models provides quantitative insights to the determinants of successful job interviews. 
With the knowledge presented in this article, we could train a system to help underprivileged youth receive feedback on job interviews that require a significant amount of social skills. 
The framework could also be expanded to help people with social difficulties, train customer service professionals, or even help medical professionals with telemedicine.


%


\ifCLASSOPTIONcompsoc
  \section*{Acknowledgments}
\else
  \section*{Acknowledgment}
\fi

The authors would like to thank Leon Weingard for helping with transcribing the audio, and Michaela Kerem for her extensive feedback.

\ifCLASSOPTIONcaptionsoff
  \newpage
\fi



\bibliographystyle{IEEEtran}
\bibliography{interview}

%








\end{document}